# Nitrogen-Vacancy Magnetometry of Edge Magnetism in $WS_2$ Flakes

*Ilja Fescenko[1,&,*] Raman Kumar,[2,&] Thitinun Gas-Osoth,[3,4,5,&] Yifei Wang,[3,4,&] Suvechhya Lamichhane,[6] Tianlin Li,[6] Adam Erickson,[3] Nina Raghavan,[6] Tom Delord,[2] Cory D. Cress,[7] Nicholas Proscia,[7] Samuel W. LaGasse,[7,8] Sy-Hwang Liou,[6] Xia Hong,[6] Jose J. Fonseca,[7] Toshu An,[4] Carlos A. Meriles,[2,*] and Abdelghani Laraoui[3,6,*]*

[1]Laser Center, University of Latvia, Jelgavas St 3, Riga, LV-1004, Latvia
[2]Department of Physics, CUNY-City College of New York, 85 St. Nicholas Terrace, New York, NY 10031, USA
[3]Department of Mechanical & Materials Engineering, University of Nebraska-Lincoln, 900 N 16th St, W342 NH, Lincoln, NE 68588, USA
[4]Japan Advanced Institute of Science and Technology, 1-1 Asahidai, Nomi, Ishikawa 923-1292, Japan
[5]Present address: School of Science, Department of Physics, University of Phayao, 19 Moo 2, Tambon Maeka, Mae Ka, Phayao 56000, Thailand
[6]Department of Physics and Astronomy and the Nebraska Center for Materials and Nanoscience, University of Nebraska-Lincoln, 855 N 16th St, Lincoln, NE 68588, USA
[7]Electronics Science and Technology Division, US Naval Research Laboratory, 4555 Overlook Ave SW, Washington, DC 20375, USA
[8]Laboratory for Physical Sciences, 8050 Greenmead Drive, College Park, MD 20740
[&]Equal contributions
[*]Corresponding Authors: iliafes@gmail.com, cmeriles@ccny.cuny.edu, alaraoui2@unl.edu

## Abstract

Two-dimensional (2D) magnets are of significant interest both as a platform for exploring novel fundamental physics and for their potential in spintronic and optoelectronic devices. Recent bulk magnetometry studies have indicated a weak ferromagnetic response in $WS_2$, and theoretical predictions suggest edge-localized magnetization in flakes with partial hydrogenation. Here, we use room-temperature wide-field quantum diamond magnetometry to image pristine and Fe-implanted $WS_2$ flakes of varying thicknesses (45–160 nm), exfoliated from bulk crystals and transferred to NV-doped diamond substrates. We observe direct evidence of edge-localized stray magnetic fields, which scale linearly with applied external magnetic field (4.4–220 mT), reaching up to ±4.7 μT. The edge signal shows a limited dependence on the flake thickness, consistent with dipolar field decay and sensing geometry. Magnetic simulations using five alternative models favor the presence of edge magnetization aligned along an axis slightly tilted from the normal to the $WS_2$ flake's plane, consistent with spin canting in antiferromagnetically coupled edge states. Our findings establish $WS_2$ as a promising platform for edge-controlled 2D spintronics.

## 1. Introduction

The recent discovery of two-dimensional (2D) magnetism[1–5] has garnered significant interest, as the ability to stack atomic layers of different magnetic materials or the same material with varying relative orientations offers countless opportunities for both fundamental and applied research. Particularly promising among these applications are spin-based technologies, such as energy-efficient memories,[6] non-volatile memristive devices,[7] and random number generators,[8] systems that have traditionally relied on dilute magnetic semiconductors. Significant progress has been made in characterizing and understanding the ferromagnetic properties of Cr- and Fe-based 2D systems,[9,10] but their metallic or insulating properties hinder seamless device integration due to challenges in electrical compatibility and interface engineering.[11]

To address these limitations, substantial efforts have been dedicated to investigating transition-metal dichalcogenides (TMDs) such as $MoSe_2$, $MoS_2$, and $VSe_2$,[12–16] where first-principles calculations predicted room-temperature ferromagnetism down to the monolayer level when doped with Fe, Co, or Mn.[9,17,18] Experimental studies have largely confirmed these predictions, demonstrating ferromagnetism in few-layer TMD flakes containing various dopants. In addition to those mentioned above,[19,20] the list has expanded to include V,[16,21] Ni,[22] Cu,[23] and Nb.[20]

Tungsten disulfide ($WS_2$), another member of the TMD family, has primarily garnered attention for optoelectronic applications for its tunable bandgap, which changes from indirect to direct when transitioning from bulk to monolayer form (~2 eV).[24,25] Recent bulk vibrating sample magnetometer (VSM) and superconducting quantum interference device (SQUID) measurements of pristine $WS_2$ nanosheets showed intrinsic room-temperature ferromagnetism,[26,27] hence extending the range of potential applications to spintronics.

The origin of ferromagnetism in $WS_2$ nanostructures remains, however, not fully understood. *Ab initio* calculations suggest the presence of spin-polarized edge states along the zigzag edges of $WS_2$ flakes, even in the absence of externally applied magnetic fields.[28] SQUID magnetometry measurements on $WS_2$ nanosheets reveal a ferromagnetic-like hysteresis, whereas bulk $WS_2$ does not exhibit such behavior.[26,27] A separate study showed ferromagnetism in Fe:$MoS_2$ but not in Fe:$WS_2$, which was attributed to the formation of deep level traps.[29]

Despite these preceding studies, the local magnetic properties of pristine or doped thin $WS_2$ flakes and sheets remain underexplored, primarily due to the lack of characterization methods featuring suitable spatial resolution and magnetic sensitivity. To circumvent this problem, the last decade has witnessed a sustained effort aimed at the use of color centers as local magnetic probes. A particularly successful example is the nitrogen-vacancy (NV) center in diamond, a paramagnetic defect whose spin can be initialized and readout optically:[30–32] NV-based optically detected magnetic resonance (ODMR) has now evolved to allow the detection of weak magnetic fields with nanometer-scale spatial resolution, thereby offering new opportunities for investigating the magnetic properties of thin films[33–37] and 2D magnets.[38–41]

A range of techniques has been employed to study magnetism in 2D materials, including magneto-optical Kerr effect (MOKE),[1,2] magnetic force microscopy (MFM),[23,42] and electron holography.[43,44] However, these methods face intrinsic limitations when applied to weak, spatially localized magnetic signals such as the edge stray fields expected from $WS_2$. MOKE lacks the sensitivity to detect stray fields in the μT range and typically integrates signals over micrometer-scale areas.[1,2] MFM, while offering high spatial resolution (< 50 nm), probes magnetic field gradients rather than absolute magnetic field values and may perturb delicate magnetic structures due to the interaction with the magnetic tip.[23,42] Electron holography requires invasive sample preparation (e.g., focused ion beam thinning) and operates under vacuum, precluding in situ magnetic field control.[43,44] In contrast, NV center-based magnetometry enables non-invasive, quantitative imaging of magnetic fields with sub-micrometer spatial resolution and nT–μT sensitivity under ambient conditions.[32,38,39,45–47] This makes NV magnetometry uniquely suited for probing edge-localized magnetic phenomena in exfoliated 2D materials.

Here, we investigate the room-temperature magnetic response of pristine and Fe implanted $WS_2$ flakes under a varying magnetic field. In our experiments, we transferred pristine and Fe implanted $WS_2$ flakes onto diamond films engineered to host an ensemble of near-surface (~6–150 nm) NV centers. NV stray-field imaging of individual flakes featuring varying thickness revealed clear magnetic signatures in both pristine and Fe-implanted $WS_2$, which allowed us to examine five different models of magnetization. [26,28] Our results provide the first direct evidence of edge-localized magnetism in $WS_2$ flakes, revealing its amplitude, orientation, and spatial distribution.

## 2. Results and Discussion

### 2.1. NV magnetometry of pristine $WS_2$ flakes

Our experiments leverage advances on the use of NV centers in diamond as local magnetic probes (see Methods for detailed discussion). The NV center is formed by a substitutional nitrogen adjacent to a vacancy and features a ground state spin triplet whose relative populations can be optically read out utilizing spin-selective intersystem crossing to a singlet manifold[48] (Figure 1a). Figure 1b shows a schematic of the NV widefield microscope.[30,31] We optically excite the NVs via a green laser (130 mW, 532 nm), and collect the NV fluorescence (650 – 800 nm) via a sCMOS camera (Supporting Information, Section S2). We position the diamond above a glass coverslip patterned with Ti /Cu thick striplines (respectively 5 nm and 1.5 µm thick) for microwave (MW) manipulation of the NV spin states.

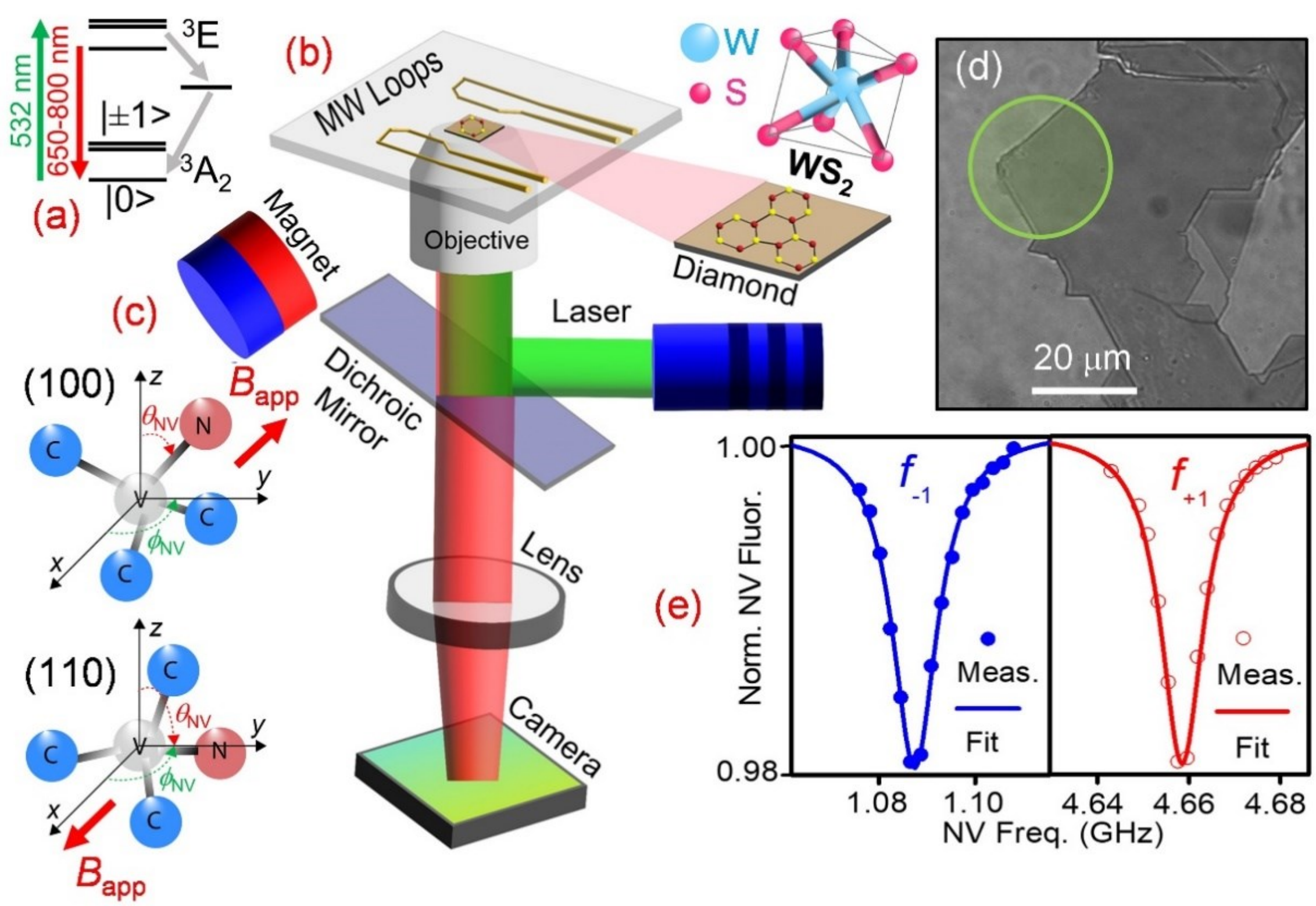

**Figure 1.** (a) Schematic of the NV center electronic energy levels. (b) ODMR widefield microscope setup. The inset of (b) illustrates the $WS_2$ unit cell. (c) Relative NV axis orientations for (100) and (110) diamonds used in this study (top and bottom, respectively). (d) Optical image of a $WS_2$ Flake 1 on a (100) diamond surface; the green circle marks the area illuminated during measurements. (e) NV ODMR spectrum in the (100) diamond at a magnetic field of 63.2 mT aligned along the NV [111] axis.

A magnetic field is applied along the [111] orientation for both (100) and (110) oriented diamonds (Figure 1c).[25,26,34] When magnetized, the $WS_2$ flake (Figure 1d) generates a stray magnetic field $B_{str}$ along the NV axis, which leads to a change in the effective magnetic field $B_{eff} = B_{app} + B_{str}$ acting on the NVs, thus altering the $f_{-1}$ ($|0\rangle \leftrightarrow |-1\rangle$) and $f_{+1}$ ($|0\rangle \leftrightarrow |+1\rangle$) NV spin transition frequencies.[30,31] Figure 1e depicts the NV ODMR resonances in the (100) diamond at $B_{app}$ = 63.2 mT, fitted with a Lorentzian to determine $f_-$ and $f_+$. We transferred three pristine and one Fe-implanted $WS_2$ flakes — thereafter referred to as Flakes 1, 2, 3, and 4, these were deposited on diamond substrates cut and polished along the (100) plane for Flakes 1 and 3, and the (110) plane for Flake 2 and Flake 4. In all diamonds, we align $B_{app}$ along the NV crystalline [111] axis corresponding to angles ($\theta_{NV}$, $\varphi_{NV}$) of (54.7°, 90°) and (90°, 0°) for the (100) and (110) planes, respectively (Figure 1c).

Figures 2(a-c) show $B_{str}$ maps from a region of pristine Flake 1 (thickness of 160 nm, measured by atomic force microscopy (AFM) in Figure S1.5, Supporting Information) for applied magnetic fields of 4.4, 39.6, and 63.2 mT, respectively. We find the NV signal can have positive or negative signs on different edges as clearly seen in Figure 2c. To evaluate the magnetic response, we make line cuts of the measured $B_{str}$ at the flake edges, as indicated by the red and blue dashed lines, and plot them separately in Figures 2d and 2e. We define the stray field strength $\Delta B$ as a peak-to-peak amplitude between the maximum and minimum in the $B_{str}$ signal.[30,31] The absolute value of $\Delta B$ increases from 0.75 ± 0.3 μT at 4.4 mT to 4.7 ± 0.3 μT at 63.2 mT. As seen in Figure 2f, the stray field amplitude grows linearly with the applied magnetic field, indicative of weakly ferromagnetic[29] or paramagnetic[30,31] behaviors, discussed below.

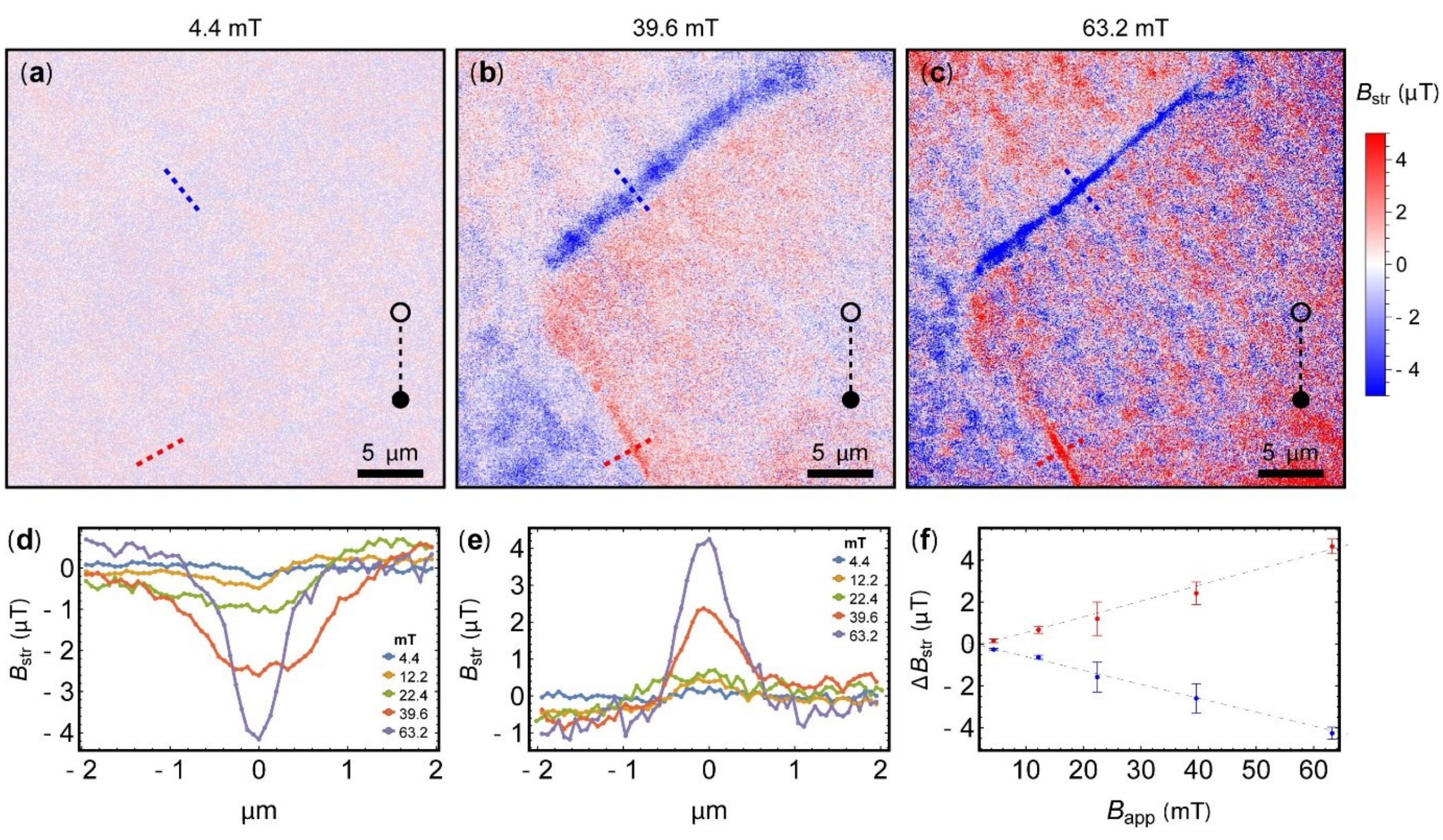

**Figure 2**. (a-c) NV $B_{str}$ maps of a $WS_2$ Flake 1 under an applied magnetic field $B_{app}$ of 4.4 mT, 39.6 mT, and 63.2 mT, respectively. The black dashed line with circles indicates the NV axis projection on the image *x*-y plane. (d-e) Line cuts of $B_{str}$ across the $WS_2$ flake edges, corresponding to the blue and red dashed lines in (a-c), revealing field variations at different applied magnetic fields. (f) Magnetic shift $\Delta B_{str}$ extracted from (d-e) as a function of $B_{app}$; the dashed lines are linear fits. The error bars in (f) represent the standard deviation of the background magnetic field in regions away from the edges.

Figures 3b and 3e display the measured $B_{str}$ from Flake 1 and Flake 2 (thickness of ~100 nm): Unlike the "absorptive" stray field profile seen for Flake 1, Flake 2 exhibits a "dispersive" shape, which is a consequence of the varying relative orientations of the NV axis and the flake magnetization at a given edge.

To quantitatively interpret these patterns, we perform three-dimensional magnetostatic modeling using finite-element analysis (COMSOL Multiphysics), see Supporting Information Section S4 and Figure S4. Figure 3c and Figure 3f show the calculated stray magnetic field in either case. These edge patterns are a result of projecting $B_{str}$ along the NV axes corresponding to each diamond orientation (i.e., along a 54.7° out-of-plane vector in the (100) diamond and in-plane for the (110) case). To investigate the nature of magnetism in $WS_2$, we consider five distinct magnetization geometries (subsequently referred to as models). Comparison of the measured data with simulations allows us to determine the most likely magnetization model, irrespective of preceding theoretical expectations. Specifically, for each model, we simulated profiles along the dashed lines shown in Figure 3b and Figure 3e, fitting their amplitudes to the measured stray field profiles. The upper rows of Figure 4a and Figure 4b schematically illustrate the geometries of the tested models. The blue arrows indicate the applied magnetic field, consistently aligned along the NV [111] axis, as depicted by a dumbbell drawing. The brown arrows represent uncompensated spins responsible for magnetization.

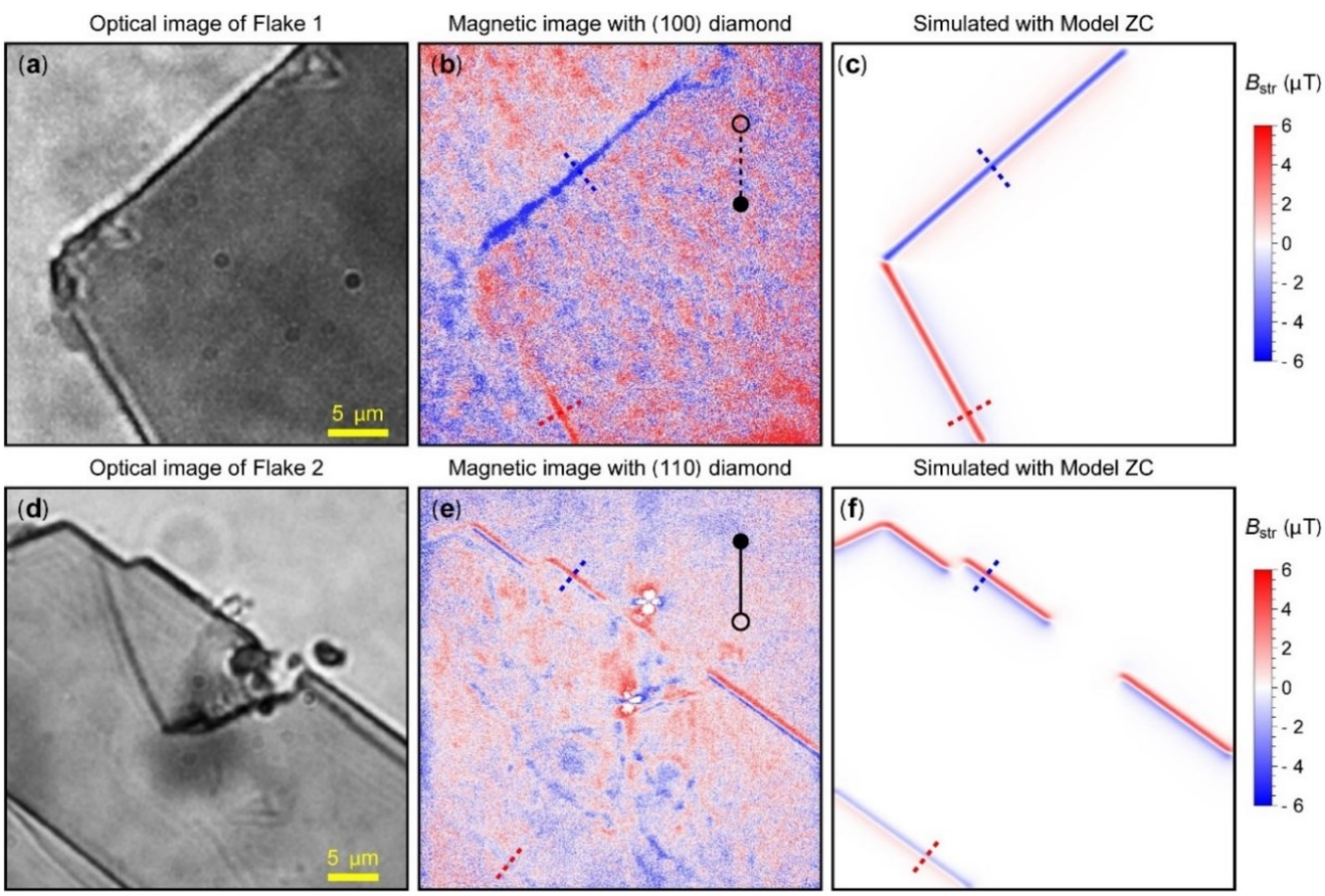

**Figure 3**: Optical images of $WS_2$ Flake 1 (a) and Flake 2 (d) on diamond substrates. Stray field magnetic images acquired at $B_{app}$ of 63.2 mT and 220 mT of Flake 1 (b) and Flake 2 (e) transferred to (100) and (110) diamond, respectively. The applied magnetic field is aligned along the diamond [111] axis. The black lines with circles represent the NV axis projection in the image plane. Simulated stray field maps of Flake 1 (c) and Flake 2 (f) using Model ZC, which assumes edge magnetization along the z-axis normal to the image plane.

Figure 4 compares the experimental NV magnetometry profile with simulated results for five different magnetic states. In Model P (Paramagnetic), we assume that the entire flake is paramagnetically magnetized in the applied field, whereas in Model PE (Paramagnetic Edge), only the edges exhibit paramagnetic behavior. We model edge magnetization by assigning a uniform volume magnetization to a bar with a cross-section of 200 nm (width) × 300 nm (height), placed along the edge of an otherwise non-magnetized flake. In Figure 3 and Figure 4, the best-fit volume magnetizations are 50 A/m for Flake 1 and 80 A/m for Flake 2. Given the applied magnetic fields of 63.2 mT and 220 mT, respectively, the higher value for Flake 2 likely corresponds to a saturated magnetization state. To estimate the magnetic moment associated with the edge, one must convert the modeled volume magnetization using the ratio between the cross-section of the model bar and the actual magnetized volume, which is approximately given by the flake height multiplied by the in-plane lattice constant. To account for the limited spatial resolution, the simulated magnetic image was convolved with a Gaussian kernel of 325 nm, effectively smoothing sharp field gradients at the bar edges. The finite cross-section of the bar has negligible influence on the resulting field profile and serves as a reasonable approximation of an edge-localized magnetization in our wide-field NV microscope. This microscope operates at diffraction-limited resolution, detecting fluorescence at wavelengths >650 nm[30,31] and thus any spin volume smaller than approximately half the detection wavelength appears spatially broadened into a diffraction-limited spot. The assumption that the magnetized region of the $WS_2$ flakes is smaller than the diffraction-limited spot is supported by the observation that the widths of the stray-field signals remain consistent across all measurements (Figures 2–5).

In Model N (Normal), we assume that uncompensated magnetic moments are oriented perpendicular to the edge faces of the flakes, forming a 90-degree angle with the side faces, while Model Z posits that the magnetic moments are aligned along the *z*-direction, normal to the imaging and flake's plane. Drawing on References [26,28] and our experimental observations, we propose that the magnetic moments could be oriented either along or opposite to the *z*-axis.

While the first three models (P, PE, and N) significantly depart from the measured data (see bottom rows of Figure 4a and Figure 4b), Model Z shows instead good agreement across both geometries. To address minor inconsistencies — such as the slightly asymmetric dispersive experimental profile in Figure 4b or the lack of asymmetry in the experimental profiles in Figure 4a — we introduce the angle between the *z*-axis and the magnetization (in the *yz* plane) as a fitting parameter. The resulting fit, shown in Figure 4 as Model ZC (Z Canting), is attributed to spin canting in antiferromagnets, where magnetic moments deviate from perfect antiparallel alignment due to external magnetic fields, Dzyaloshinskii–Moriya interactions, or crystal anisotropy, resulting in a small net magnetic moment.[49]

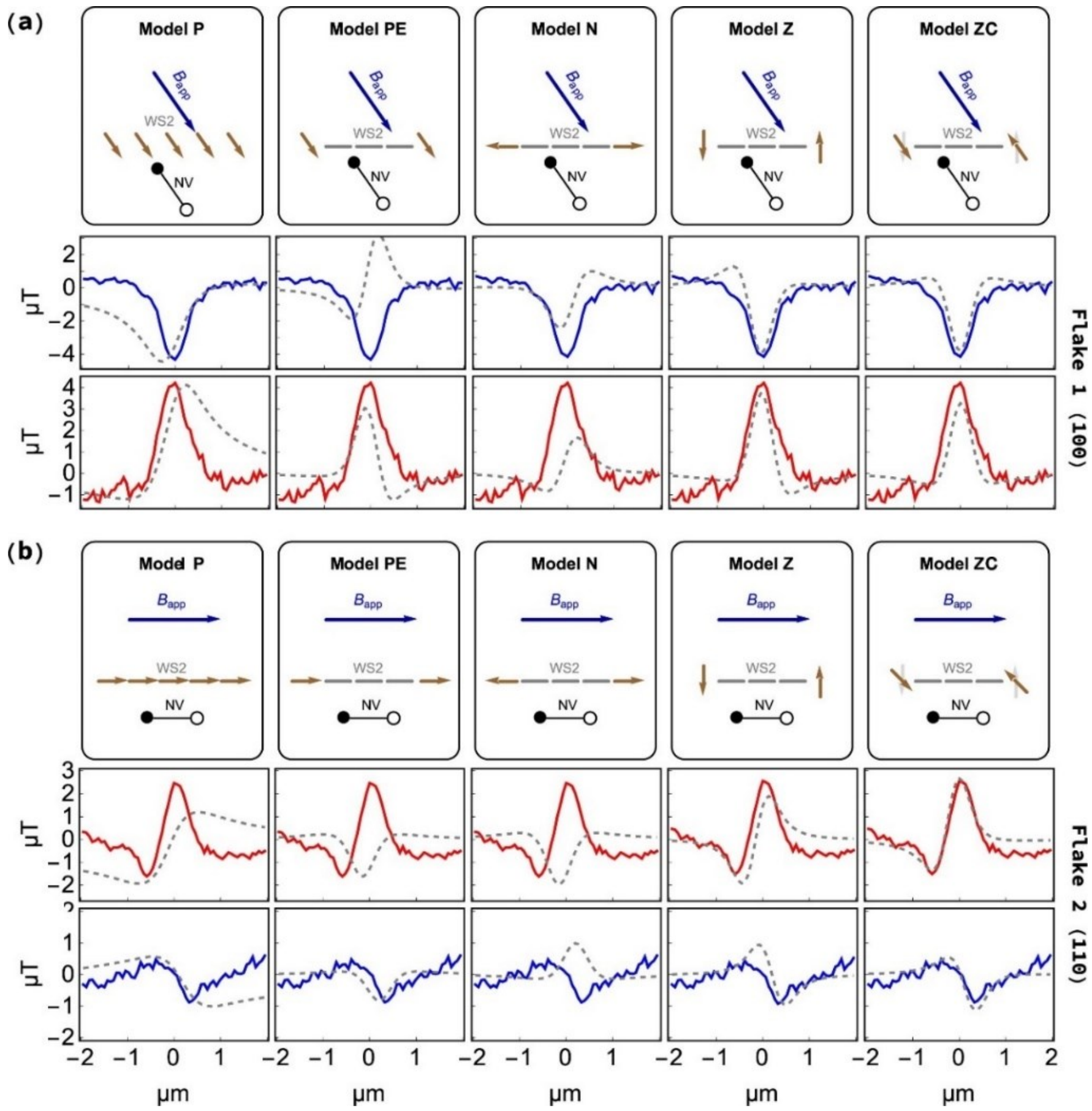

**Figure 4**: Comparison of simulated and experimental NV magnetometry profiles for $WS_2$ Flake 1 (a) and Flake 2 (b) with different edge magnetization configurations. The top (a, Flake 1) and bottom (b, Flake 2) panels illustrate simulations for distinct magnetization geometries of $WS_2$ edges under an applied magnetic field $B_{app}$. The experimental data (solid red and blue lines) correspond to NV magnetometry profiles measured along opposite edges of $WS_2$ flakes. Simulated profiles (dashed gray lines) in (a-b) are overlaid for direct comparison.

The observed positive and negative stray-field features consistently exhibit similar amplitudes, which does not fully align with theoretical predictions[29] suggesting that different atomic species in monolayer $WS_2$ (i.e., S and W atoms) contribute unequally to the local magnetic moment. However, when comparing different flakes, the amplitudes of the magnetic signals show a rather small dependence on flake thickness. Pristine Flake 3 (thickness of 45 nm from AFM in Figure S1.5) exhibits a smaller magnetic signal compared to Flake 1 (thickness of 160 nm) under the same applied field of 63.2 mT (see Figure S3.3). The reduction in signal amplitude is only about a factor of two, while the difference in spatial width is masked by the diffraction-limited resolution. This relatively modest amplitude change is expected: while a thicker edge contains more dipoles, the stray magnetic field decays as $1/r^3$, meaning that dipoles farther from the NV plane contribute less. For flakes of 45 nm and 160 nm thickness, simulations predict a difference in stray field amplitude of less than 2%. Therefore, the observed measured difference is most likely due to a slightly larger NV–sample standoff for Flake 3 compared to Flake 1. We note that we used a hexagonal boron nitride (hBN) flake (thickness of ~100 nm) to cap the thin $WS_2$ Flake 3 and prevent contamination/oxidation. As shown in Figure S3.3a in Supporting Information, this led to an increase of fluorescence (~2 times higher than the NV fluorescence) from unpolarized green excitation at the edges between hBN and $WS_2$ flakes, that may be explained by the activation of other quantum emitters as a result of carrier trapping in deformation potential wells localized near the edges between hBN and $WS_2$,[50] or by waveguiding effects induced by hBN.[51] This extra fluorescence can affect the NV ODMR contrast in Flake 3 in comparison to Flake 1 (no hBN) at the same laser power (130 mW) and averaging time (10 seconds per frame).

### 2.2. NV magnetometry of Fe-implanted $WS_2$ flakes

We have also conducted NV measurements on selected Fe-implanted (energy of 30 eV, dose of $10^{13}$ – $10^{14}$ $Fe^+/cm^2$) $WS_2$ flakes transferred onto (110) diamond, as shown in Figure 5a. More information about $Fe^+$ beam implantation is provided in Supporting Information Section S1. Figure 5b and Figure 5c respectively present the measured and simulated $B_{str}$ maps at an applied field of 182 mT, corresponding to a region highlighted by circles in Figure S1.3. The profiles along the dashed lines were fitted to simulated profiles using Model ZC, as depicted in Figure 5c. The width and shape of these profiles (Figure 5d) resemble those of pristine flakes (similar Flake 2 transferred into (110) diamond), though no evidence of uniform magnetization due to Fe implantation was observed across the flake. This suggests that magnetism in $WS_2$ flakes originates most likely from the flake edges[26,28] or other local defects present in the material,[52] rather than the Fe doping.[29] This behavior is confirmed from previous bulk measurements on $WS_2$ powder, in which the magnetization is enhanced by increasing the annealing temperature up to 700 °C and therefore increasing the edge defects density, confirmed from X-ray photoelectron and electron paramagnetic resonance spectroscopies.[52] Nevertheless, the low energy $Fe^+$ shallow (~1 – 3 nm) implant without annealing, may not be enough to activate Fe centers within the $WS_2$ flakes. Future studies on flakes exposed to varying dose and implantation energy of $Fe^+$ ions may show a measurable response in the Fe-implanted $WS_2$ flakes.

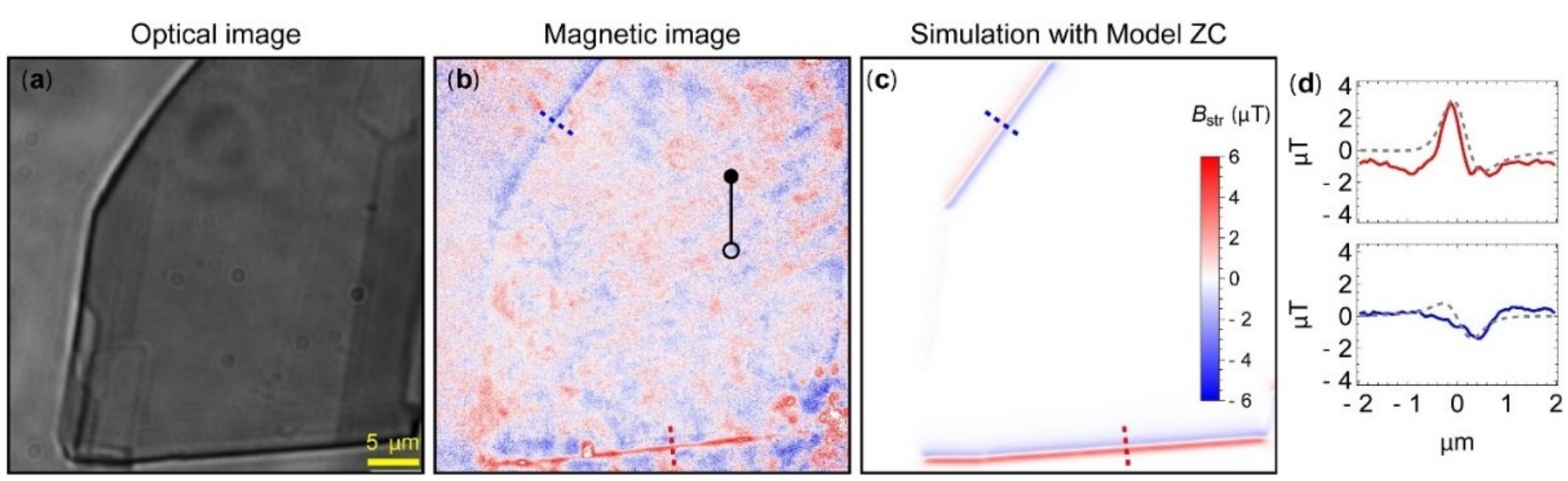

**Figure 5.** Optical image (a), measured (b), and simulated stray field map using Model ZC (c) which assumes edge magnetization along the z-axis normal to the image plane with weak paramagnetic behavior. The experimental data (solid red in d and blue lines in e) correspond to NV magnetometry profiles measured along opposite edges of $WS_2$ flake in (b). Simulated profiles (dashed gray lines) in (d) are overlaid for direct comparison.

## 3. Conclusion

To conclude, we have employed wide-field NV magnetometry to map the room-temperature magnetic stray fields of pristine and Fe-implanted $WS_2$ flakes (45–160 nm thick) transferred onto (100) and (110) diamond substrates. Our measurements reveal distinct magnetic signatures localized at the flake edges, with stray fields up to ±4.7 µT that scale linearly with the applied magnetic field (4.4–220 mT). By comparing experimental data with finite-element simulations across five magnetization models, we have found that the observed profiles can be best described by edge-localized magnetization slightly tilted from the *z*-axis. This tilt suggests a complex interplay between intrinsic edge magnetism and the applied magnetic field, partially deviating from theoretical predictions of purely *z*-aligned spin-polarized edge states.

Notably, Fe-implanted flakes exhibit edge magnetism similar to pristine flakes, with no evidence of uniform magnetization induced by doping, supporting the hypothesis that magnetism in $WS_2$ originates primarily from zigzag edge defects rather than dopants. The modest dependence of stray field amplitude on flake thickness (e.g., ~2-fold reduction from 160 nm to 45 nm) aligns with the $1/r^3$ decay of dipole fields, where thicker edges contribute more dipoles, but distant ones have diminished impact at the NV sensing plane (~6–150 nm standoff).

These findings provide direct evidence of edge-localized ferromagnetism in $WS_2$ flakes at room temperature, advancing our understanding of 2D magnetism in transition metal dichalcogenides. The ability to probe nanoscale magnetic features with NV magnetometry highlights its potential as a tool for characterizing 2D magnets with high sensitivity and spatial resolution.[32,42] However, the limited spatial resolution (~325 nm) of wide-field NV microscopy obscures finer details of edge magnetization, and the exact nature of the observed tilt in the ZC model remains unclear, potentially reflecting external or internal field effects.

Future studies could leverage NV-scanning probe microscopy[32,38,53] (~50 nm resolution) with NV ensembles to resolve sub-edge magnetic structures with weak magnetic stray fields and exfoliate monolayer $WS_2$ flakes to explore ferromagnetic order at the atomic limit. Such efforts could enable $WS_2$-based spintronic devices,[54] such as magnetic tunnel junctions or spin valves, by harnessing edge magnetism. Additionally, combining NV magnetometry with complementary techniques (e.g., MOKE)[32,42] could clarify the interplay between edge and bulk contributions, further elucidating the mechanisms of ferromagnetism in $WS_2$ and related TMDs. In a recent theory study, it was found that the edge magnetism in TMDs depends highly on the edge filling of electronic states.[55] Therefore, a modulation (e.g., electrical) of the edge magnetic phases could be experimentally observed in gated TMD samples including $WS_2$.

## 4. Methods

*Samples preparation*: $WS_2$ flakes with a thickness of 45–160 nm were prepared by mechanical exfoliation from a bulk $WS_2$ crystal (HQ Graphene) onto a 100 nm $SiO_2$/Si substrate with pre-patterned etched marks, see Supporting Information, Figure S1.1. After thermal annealing, we used shadow masks and Fe implantation to create doped areas in some of the flakes (see Supporting Information Section S1 for further details). Then, select $WS_2$ flakes were picked up from $SiO_2$/Si and Au/Ti-coated $SiO_2$/Si substrates and transferred onto NV doped diamond using high-temperature dry transfer[56,57] with polypropylene carbonate (PPC) stamps. Thicker $WS_2$ flakes were transferred directly using PPC (Figure S1.4(a-c)), while we added a protective top layer of hBN on PPC for thinner flakes (Figure S1.4(d-f)). To release the flakes onto the diamond, the PPC stamp was brought into contact with the substrate at 60 °C; the samples were then heated to 100 °C to fully melt the PPC, allowing the flakes to detach. The flakes' thickness is retrieved from atomic force microscopy and by correlating it with the flakes color using optical microscopy measurements, see Figure S1.1b.

*NV Magnetometry*: In this study, we used $^{15}N$ implantation[31,34,58] to create ~6-nm[34,58] and ~150-nm[30,31] thick NV sensing layers into two $2 \times 1 \times 0.08$ mm$^3$ electronic grade (100) and (110) diamonds (see Supporting Information Section S2 and Figure S2(a-b) for more details on NV creation). For NV measurements, we first locate the $WS_2$ flake of interest with the help of a sCMOS camera and then record the ODMR signal by steering the NV fluorescence into an avalanche photodetector connected to a Yokogawa oscilloscope. To map the stray field emanating from the $WS_2$ flakes, we collect the NV fluorescence at each MW frequency by the sCMOS camera (for a total of 16 frames per spin transition), we then evaluate the ODMR dips at each pixel using an amplitude-weighted mean method,[59] as shown in Figure 1e. We calculate $B_{str}$ from the formula:[30,31]

$$B_{str} = (f_+ \pm f_-)/2\gamma_{NV} - B_{app} \quad (1)$$

where $\gamma_{NV}$ = 28 GHz/T is the gyromagnetic ratio of the NV electron spin, and we use the minus (plus) sign for $B_{app} < 102.5$ mT ($B_{app} > 102.5$ mT), take into account the ground-state spin level anti-crossing. This strategy allows us to remove any background signal coming from local strain and thermal effects leaving only the stray magnetic field produced by the $WS_2$ flakes.[30,31]

**Acknowledgements:** I.F. acknowledges support from the Latvian Quantum Initiative under European Union Recovery and Resilience Facility project no. (2.3.1.1.i.0/1/22/I/CFLA/001). A.L., S.-H.L., and X.H. acknowledge the National Science Foundation (NSF) through EPSCoR RII Track-1: Emergent Quantum Materials and Technologies (EQUATE) Award OIA-2044049. A.L. and X.H. acknowledge the University of Nebraska-Lincoln (UNL) Grand Challenges catalyst award entitled “Quantum Approaches addressing Global Threats”. A.L. acknowledges additional support from NSF Award 2328822. C.A.M. acknowledge support from NSF Award 2203904 and from the Department of Defense Award DOD-W911NF-25-1-0134. The research done at UNL was performed in part in the Nebraska Nanoscale Facility: National Nanotechnology Coordinated Infrastructure and the Nebraska Center for Materials and Nanoscience (and/or NERCF), supported by NSF Award 2025298. R.K., T.D., and C.A.M. acknowledge access to the facilities and research infrastructure of the NSF CREST IDEALS Award 2112550.

**Conflict of Interest:** The authors declare no conflict of interest.

**Author Contributions:** I.F., J.J.F., C.A.M., and A.L. designed the experiment and supervised the project. T.G.-O., Y.W., T.A., and A.L performed NV measurements and analyzed the data, with assistance from S.L., S.-H.L., and I.F. C.D.C, N.P., S.W.L., and J.J.F. synthesized the $WS_2$ flakes, conducted optical and atomic force microscopy characterization, exfoliated the $WS_2$ flakes onto a $SiO_2$/Si substrate, and performed $Fe^+$ implantation on selected $WS_2$ flakes. R.K. conducted COMSOL simulations of magnetic stray fields with assistance of T.D. T.L., N.R., and X.H. transferred $WS_2$ flakes onto the diamond using PPC and/or hBN methods. A.E. conducted additional atomic force microscopy characterization on select $WS_2$ flakes. I.F. performed the fitting using distinct magnetization models. I.F., R.K., C.A.M., and A.L. wrote the manuscript, with contributions and feedback from all authors.

**Data Availability Statement:** The data that support the findings of this study are available from the corresponding author upon reasonable request.

**Keywords:** nitrogen-vacancy, magnetometry, $WS_2$, transition metal dichalcogenide, magnetism.

# Supporting Information

## S1. Synthesis and characterization of $WS_2$ (pristine and Fe-implanted) flakes

For our experiments, we used a bulk tungsten disulfide ($WS_2$) single crystal purchased from HQ Graphene. We generated flakes via physical exfoliation and transfer. The substrate was kept at 80°C during the process to improve $WS_2$ flake transfer. Desired flake thicknesses ranging from 8 – 340 nm were preselected via optical contrast and atomic force microscopy (AFM), see Figure S1.1(b). Following flake transfer, samples were annealed at 300°C for 1hr in $H_2$/Ar environment to improve surface cleanness. Selected $WS_2$ flakes were identified optically, with thickness characterized by atomic force microscopy (AFM, Bruker Dimension FastSCAN), see Figure S1.1.

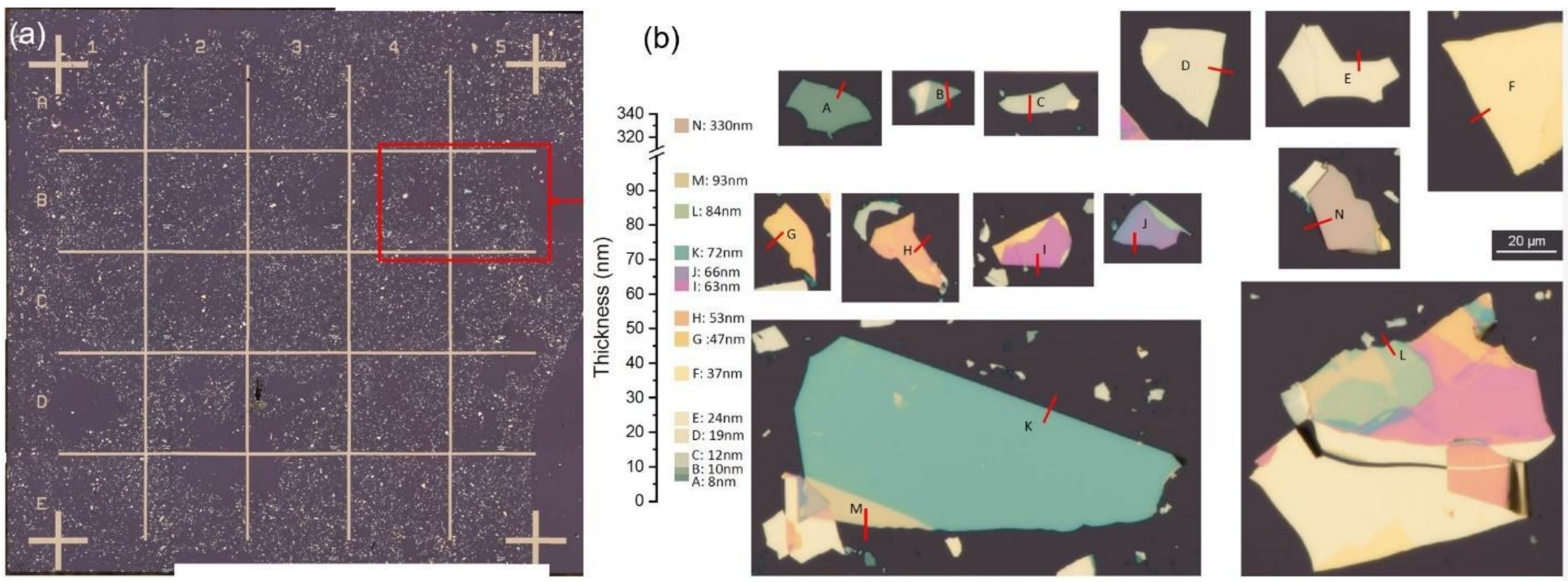

**Figure S1.1.** (a) Optical image of $WS_2$ flake exfoliated on a marked $SiO_2$/Si substrate. (b) Optical and AFM characterization of selected isolated $WS_2$ pristine flakes in the red rectangle area in (a). The color of the flake changes based on its thickness in the range of 8 – 340 nm.

### *Synthesis of Fe-implanted $WS_2$ flakes*

Fe-implanted $WS_2$ samples were fabricated by a two-step mechanical exfoliation process of $WS_2$ onto a gold coated (Au (50 nm)/Ti (5 nm)) substrate with pre-patterned trenches and holes (~200 nm deep), see Figure S1.3. First, a bulk $WS_2$ piece is cleaved obtaining a flat and pristine surface, which rapidly gets pressed and thermal released onto the gold coated substrate. Mild annealing (150°C for 15 mins) is used to improve adhesion between the $WS_2$ and the gold surface. Second, the bulk $WS_2$ piece is removed with commercial low-residue tape, obtaining mainly mono- and few-layer $WS_2$ suspended over the trenches and holes. Suspended monolayer membranes were rapidly identified by fluorescence microscopy. Thicker suspended membranes were identified with optical contrast and AFM. Carbon tape conductive adhesive was placed on either side of the sample to separate the sample from the mask and to ensure proper charge dissipation of the ion's incident with the mask.

Hyperthermal Fe ions were implanted into some of the studied $WS_2$ flakes using a custom implanter[60] based on a Colutron Ion Gun Model G-2-D with decelerator model 450 for use with ultralow energy ions down to 1 eV. A DC Ion Source with solid charge holder containing $FeCl_2$ as the Fe ion source precursor was used. Some of the studied $WS_2$ flakes were exposed to 4 nA (~1.2 nA/cm$^2$) 30 eV $Fe^+$ beams to a total Fe dose ranging from $1\times10^{13}$ to $1\times10^{14}$ $Fe^+$/cm$^2$. From Stopping and Range of Ions in Matter (SRIM) Monte Carlo simulations, the 30 eV $Fe^+$ range is 0.6 nm with a 0.3 nm struggle for a total range of ~0.9 nm (Figure S1.2a) and the 100 eV $Fe^{+}$ range is 0.9 nm with a 0.4 nm struggle for a total range of ~1.3 nm (Figure S1.2b). However, the maximum range is still greater than the projected range with a straggle of ~1.5 nm for 30 eV (see Figure S1.2a) and ~2.5 nm for 100 eV (see Figure S1.2b). The atom distributions for 30 and 100 eV implantation in Figure S1.2 show both W and S recoils with most of the W recoils are at the surface.

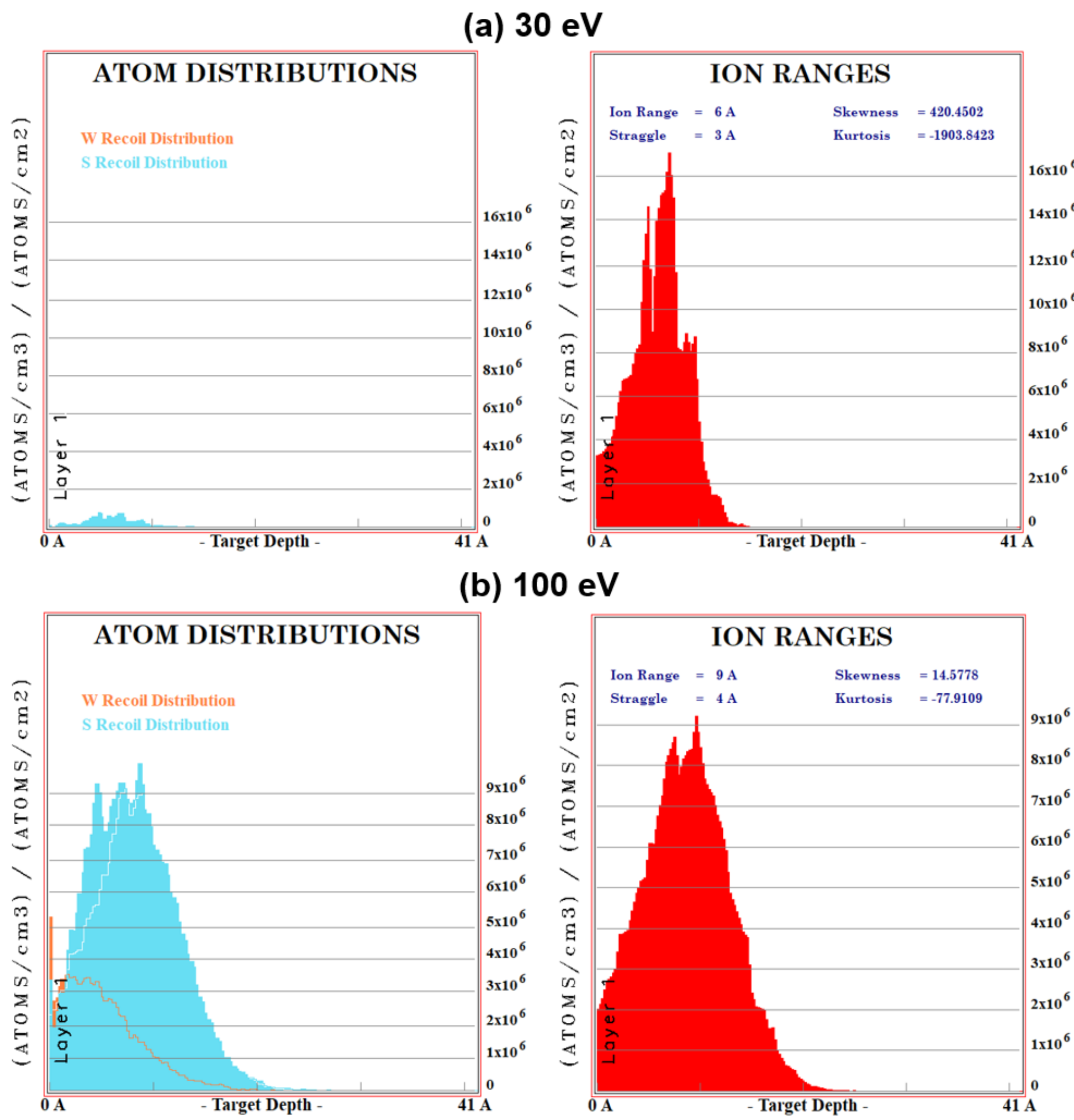

**Figure S.1.2.** SRIM calculation of $Fe^{+}$ implanted into $WS_2$ substrate with a density of 7.5 $g/cm^3$ with detailed collision cascades (atom distributions and ion ranges) for an implantation energy of 30 eV (a) and 100 eV (b).

To achieve regions with and without $Fe^{+}$ exposure on the same flake, a metal shadow mask composed of 127 μm holes on a hexagonal grid with a pitch of 220 μm was placed over the surface of the sample (Figure S1.3). Optical images were taken with the shadow mask present to determine the exposed regions of the samples based on the numbered alignment marks on the sample. Additionally, large area Raman maps (not shown here) were performed following implantation. These maps showed negligible change in the intensity for the ion implant doses imparted, yet the Raman modes have been shown to be more robust in transition metal dichalcogenides (TMD) films than photoluminescence spectroscopy at low ion doses.[61]

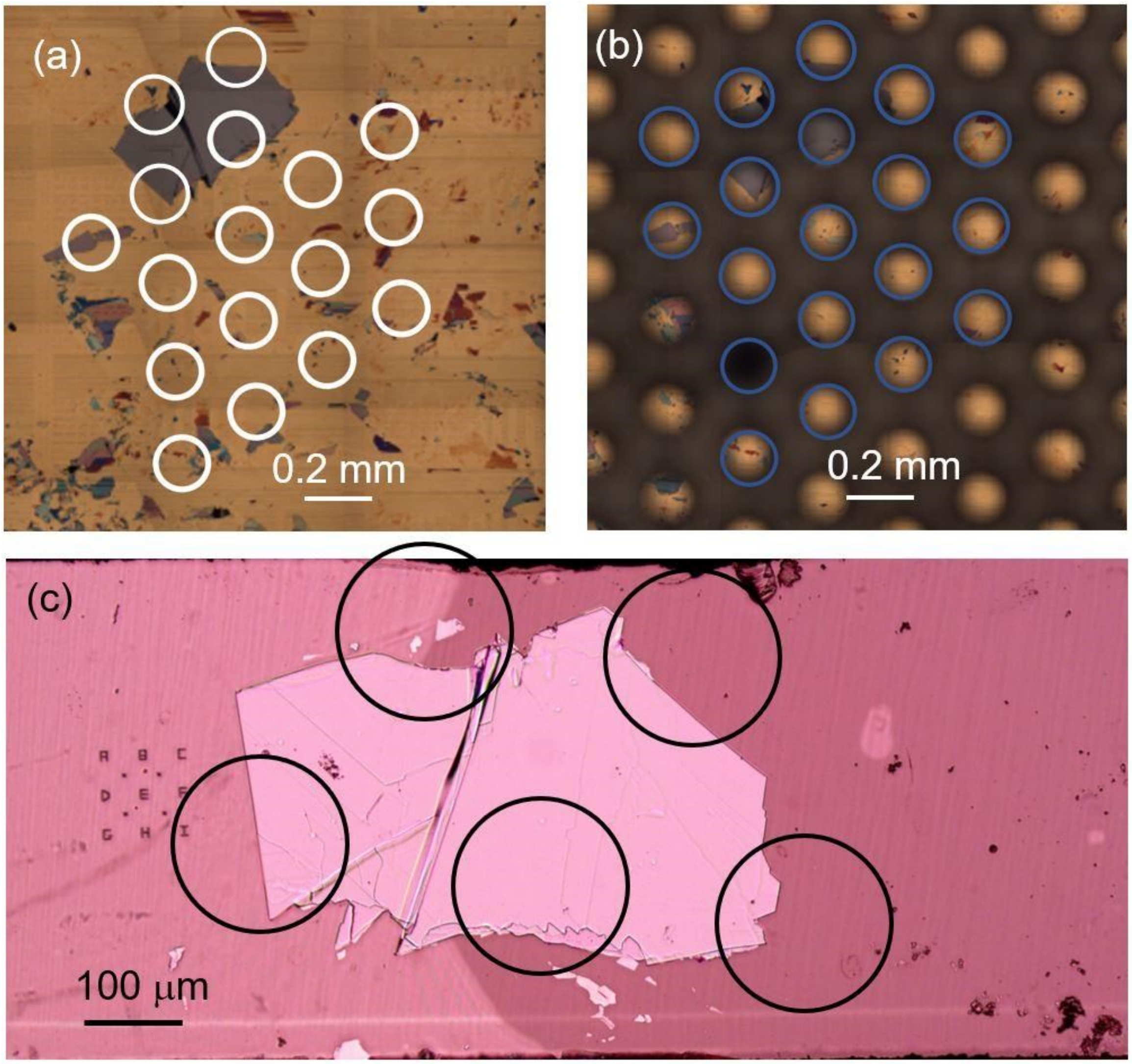

**Figure S1.3.** Optical image of a selected region with $WS_2$ fakes before (a) and after (b) depositing mask for ion beam implantation of $Fe^+$ ions. The regions highlighted with circles show the places where $Fe^+$ ions are being implanted. (c) Optical image of a pristine $WS_2$ flake transferred into a (110) diamond substance for NV measurements reported in the main manuscript and Supporting Information Section S3.

***Transfer of $WS_2$ flakes into diamond substrates***

$WS_2$ (pristine and/or Fe-implanted) flakes of interest were identified optically using standard optical color microscope[62,63] and transferred to diamond substrates (see Figures S1.3c, S1.4c, and S1.4f) using polypropylene carbonate (PPC) method (described in the main text).[64,56] After transfer, any remaining PPC residue was cleaned off using anisole. We used hexagonal boron nitride (hBN) as a capping flake (thickness ~ 100 nm) on selected $WS_2$ flakes (see Figure S1.4(d-f) to prevent contamination during the transfer.[65] Then, we resorted to nitrogen-vacancy (NV) magnetometry to image the magnetic stray field produced from pristine and Fe-implanted $WS_2$ flakes (thickness of 45 – 160 nm).

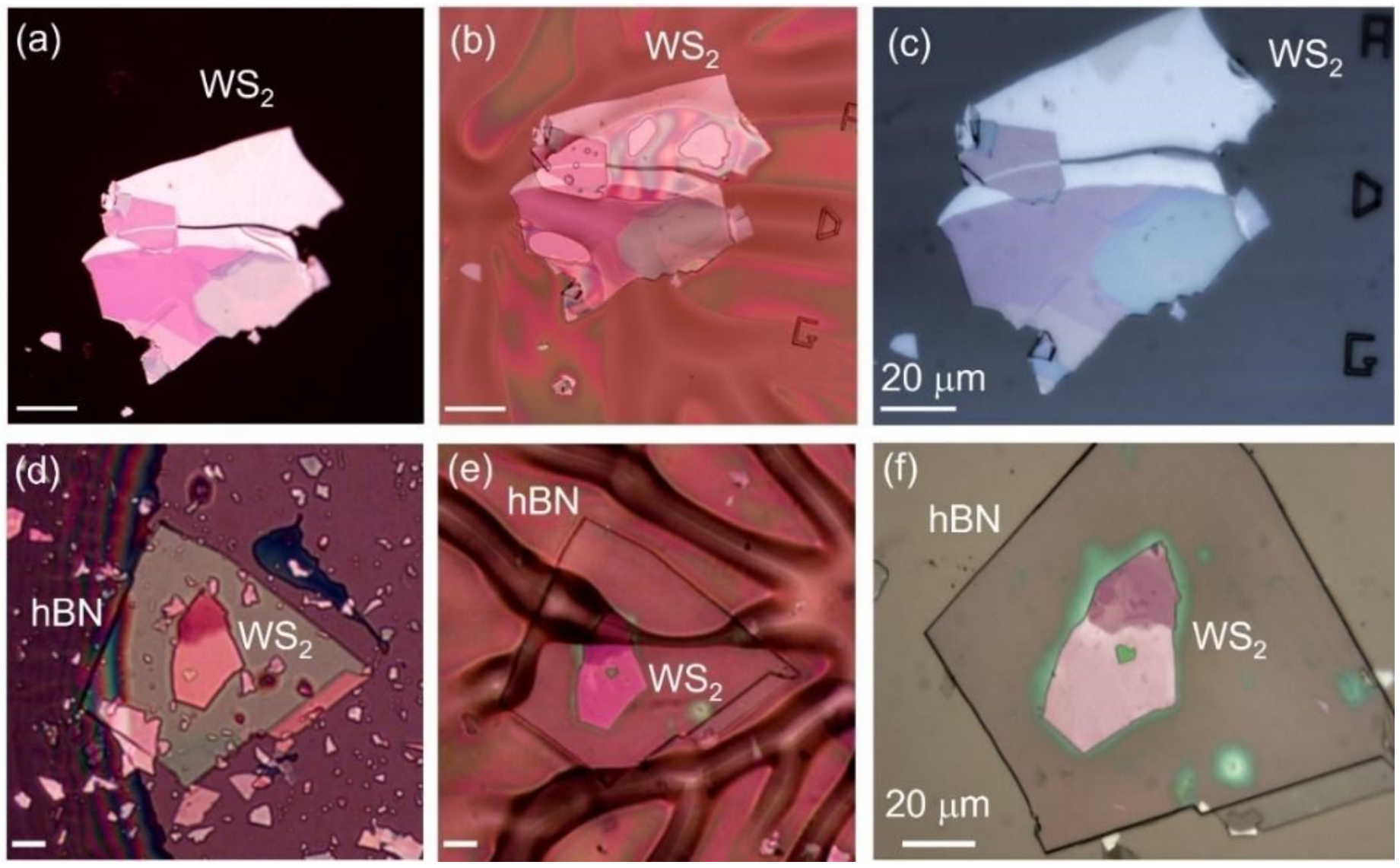

**Figure S1.4.** Optical images of a $WS_2$ flake on (a) a PPC stamp, (b) after being transferred onto the diamond substrate, and (c) after being cleaned with anisole. Optical image of a $WS_2$ flake (d) being transferred using a top layer of hBN on PPC, (e) after being transferred onto the diamond substrate, and (f) after being cleaned.

### *AFM measurements on $WS_2$ flakes transferred into diamond substrates*

To measure the thickness of the $WS_2$ flakes transferred into the diamond substrates, we performed AFM measurements by using commercial AFM from Innova from Bruker.[63] Figure S1.5a and Figure S1.5c show AFM maps of the $WS_2$ flakes transferred to (100) diamond. The flake thickness is determined from the step height in the line-cut profiles at the flake edge, see Figures S1.5b and S1.5d. The thickness of Flake 1 is ~160 nm, Flake 2 is ~100 nm, and Flake 3 is ~ 45 nm, respectively.

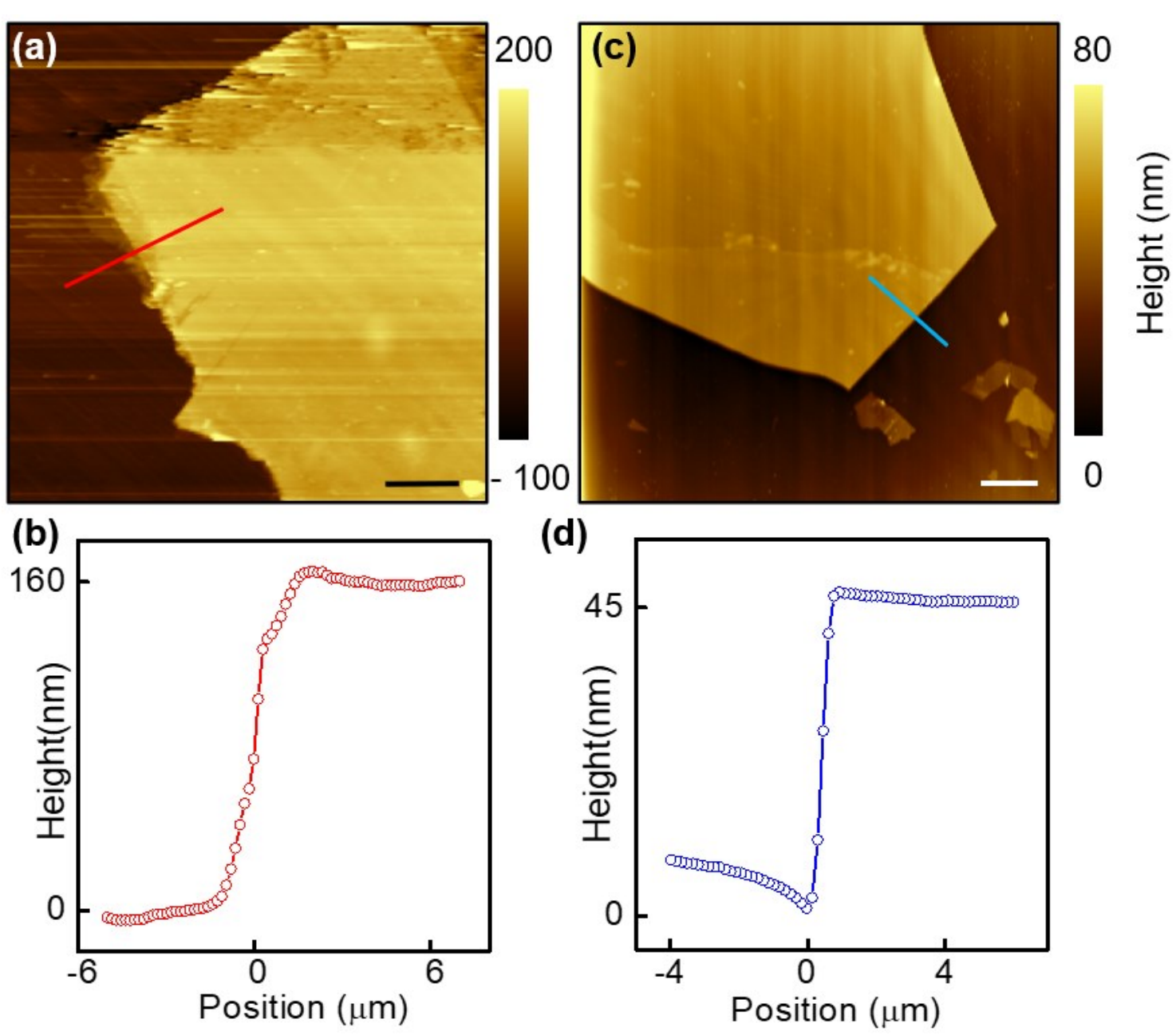

**Figure S1.5.** AFM image of $WS_2$ Flake 1 (a) and Flake 3 (c). (b) and (d) are the profiles of the flakes in (a) and (c) showing a height of ~160 nm and ~45 nm, respectively.

## S2. Experimental Setup: optical detected magnetic resonance

### *Creation of NV centers in diamond*

We used 2 mm × 2 mm × 0.5 mm type-IIa electronic grade (EL) (100) diamond (Element6) substrate with a nitrogen concentration < 5 ppb. The diamond is cut and polished along (100) and (110) planes at Delaware Diamond Knives Inc to 2 mm × 1 mm × 0.08 mm membranes. The diamond is then implanted at CuttingEdge Ions LLC with $^{15}N$ ions. The (100) EL diamond is implanted with $^{15}N$ ions at an energy of 4 keV and a dose of $1 \times 10^{13}$, similar to references [34,58]. The (110) diamond is implanted with $^{15}N$ at four energies of 10, 20, 35, and 50 keV and four doses ($1 \times 10^{13}$, $2 \times 10^{13}$, $3.5 \times 10^{13}$, and $5 \times 10^{13}$ $cm^{-2}$) respectively to create a uniform layer of NVs near the diamond surface.[30,31] We used SRIM simulations to estimate the vacancy distribution depth profile in the diamond substrate and found a uniform distribution of vacancies within ~6 nm beneath (100) diamond (Figure S2a) and ~150 nm beneath (110) diamond (Figure S2b) facing the $^{15}N$ ion source. After the implantation, the diamonds were annealed in a vacuum (pressure ~$10^{-6}$ torr) at a temperature of 800 °C for 4 hours and 1100 °C for two hours and then cleaned for four hours in a 1:1:1 mixture of nitric, sulfuric, and perchloric acid at 200 °C to remove graphite reside at the surface.[31,34,58] This process led in ~6-nm and ~150-nm NV layer near the surface with a density of ~1 ppm for (100)[34,58] and ~10 ppm for (110)[31] oriented diamond, respectively.

### *Optical magnetic resonance wide-field microscope.*

To study the magnetic properties of $WS_2$ flakes, we used home-built wide-field optically detected magnetic resonance (ODMR) microscope, Figure S2(c-d). The ODMR microscope, described in detail in Reference [9], comprises a high NA (= 1.25) 100x oil immersion Nikon objective to focus the green laser (532 nm) with a laser of 130 mW, a dichroic mirror (DM, Semrock model FF560-FDi01) and a single-band bandpass filter (Semrock FF01-731-137) to collect the NV fluorescence (650 – 800 nm) and focus it on the sCMOS camera (Hamamatsu, ORCA-Flash4.0 V3) using a tube lens (Thorlabs, TTL200). To manipulate the $|m_s = 0\rangle \leftrightarrow |m_s = -1\rangle$ ($f_-$) and $|m_s = 0\rangle \leftrightarrow |m_s = +1\rangle$ ($f_+$) NV spin transitions, a microwave (MW) is provided by a signal generator (SRS: Stanford Research Systems, SG384) with two outputs for the main (DC-4 GHz) and double-frequency harmonic (4-8 GHz). It is then amplified by two MW amplifiers (Mini-Circuits ZHL-16W-43-s+ and ZVE-3W-83+) to achieve a MW power in the range of 100 mW – 10 W. The MW is then injected through Cu (1.5 µm)/Ti (5 nm) MW loops (width of 20 µm) patterned on a glass coverslip (thickness of 100 µm).

Permanent magnets (KJ magnetics, DX8X8-N52) were used to provide an external magnetic field, applied along the [111] direction in (100) and (110) diamonds. Figures S2c shows a picture of the ODMR wide-field microscope with a magnetic field of up to 100 mT along the [111] direction in (100) diamond. For (110) diamonds (Figure S2d), the magnetic field is contained within the plane of the diamond substrate and can reach an amplitude of 400 mT. To prevent from a strong field gradient across the wide-field fluorescence image 600 pixel × 600 pixel (up to 39 µm × 39 µm), we used two permanent magnets (DX8X8-N52) connected through stainless non-magnetic steel rods, see Figure S2d.[31] To align the magnetic field along the NV [111] axis, a flip mirror is used to reflect the NV fluorescence and focus it with a lens (focal length of 30 mm) on an avalanche photon detector (APD, Thorlabs APD410A), connected to a Yokogawa oscilloscope (DL9041L). By monitoring the ODMR peaks in the oscilloscope along the [111] direction we can align the applied magnetic field very accurately. After the magnetic field alignment, the mirror is flipped down and the NVs fluorescence is imaged by the sCMOS camera, and each MW frequency is swept along the $f_-$ and $f_+$ NV spin transitions. More details for such measurements are provided in references [31,58].

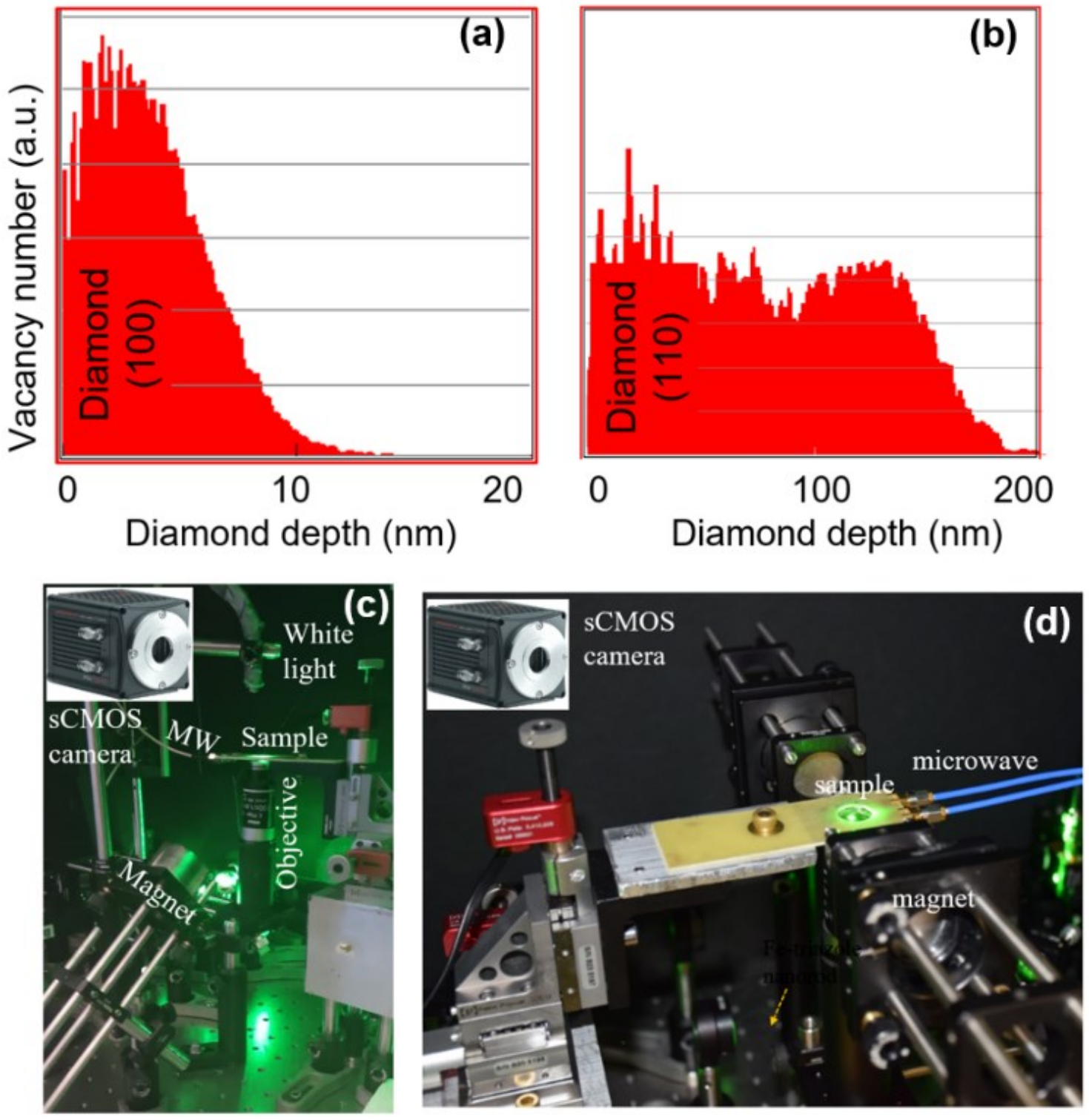

**Figure S2.** SRIM vacancy-depth profile for $^{15}N^{+}$ implantation in (100) diamond (a) and (110) diamond (b), corresponding to an NV layer of ~6 nm and ~150 nm, respectively. ODMR microscope integrated with a wide field sCMOS camera (upper inset) with magnetic field aligned along [111] in (100) diamond (c) and (110) diamond (d).

### S3. Additional NV magnetometry measurements of $WS_2$ flakes

Figure S3.1a and Figure S3.1b show the NV normalized fluorescence and measured magnetic stray-field ($B_{str}$) maps at different values (4.4 – 63.1 mT) of the applied magnetic field $B_{app}$, respectively, for the pristine $WS_2$ Flake 1 (thickness of ~160 nm) transferred into (100) diamond and studied in the main text.

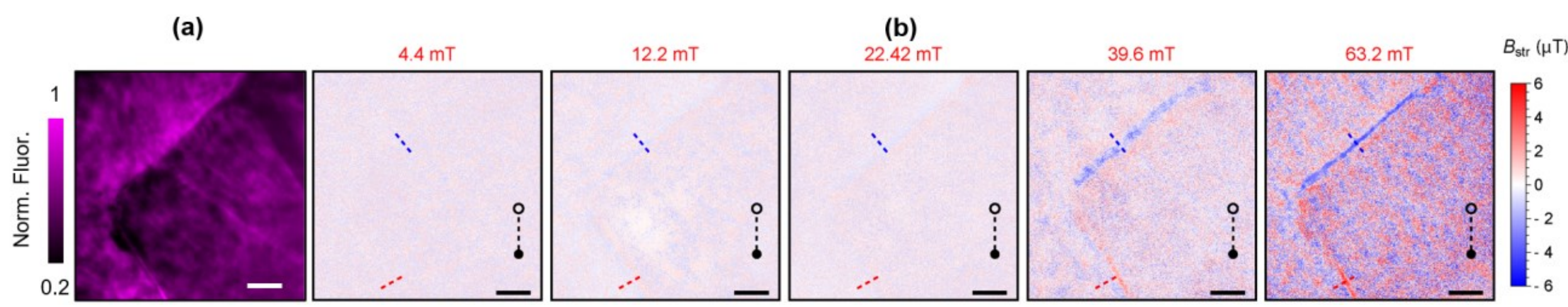

**Figure S3.1.** NV normalized fluorescence (a) and measured $B_{str}$ (b) maps at different values of $B_{app}$ (4.4 – 63.1 mT) for $WS_2$ Flake 1 transferred on (100) diamond. The scale bar in (a) and (b) is 5 μm.

Figure S3.2a and Figure S3.2b show the NV normalized fluorescence and measured $B_{str}$ maps at different values (5.6 – 220 mT) of $B_{app}$, respectively, for pristine $WS_2$ Flake 2 (thickness of ~100 nm) transferred into (110) diamond and studied in the main text.

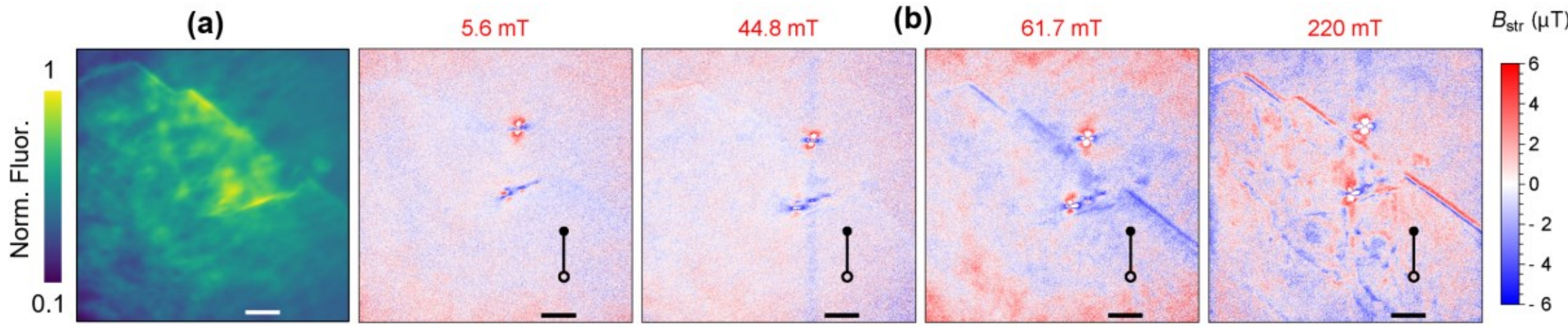

**Figure S3.2.** NV normalized fluorescence (a) and measured $B_{str}$ (b) maps at different values of $B_{app}$ (5.6 – 220 mT) for $WS_2$ Flake 2 transferred on (110) diamond. The scale bar in (a) and (b) is 5 μm.

Figure S3.3a and Figure S3.3b show the NV normalized fluorescence and measured $B_{str}$ maps at different values (5.5 – 63.2 mT) of $B_{app}$, respectively, for pristine $WS_2$ Flake 3 (thickness of ~45 nm) transferred into (100) diamond and discussed in the main text.

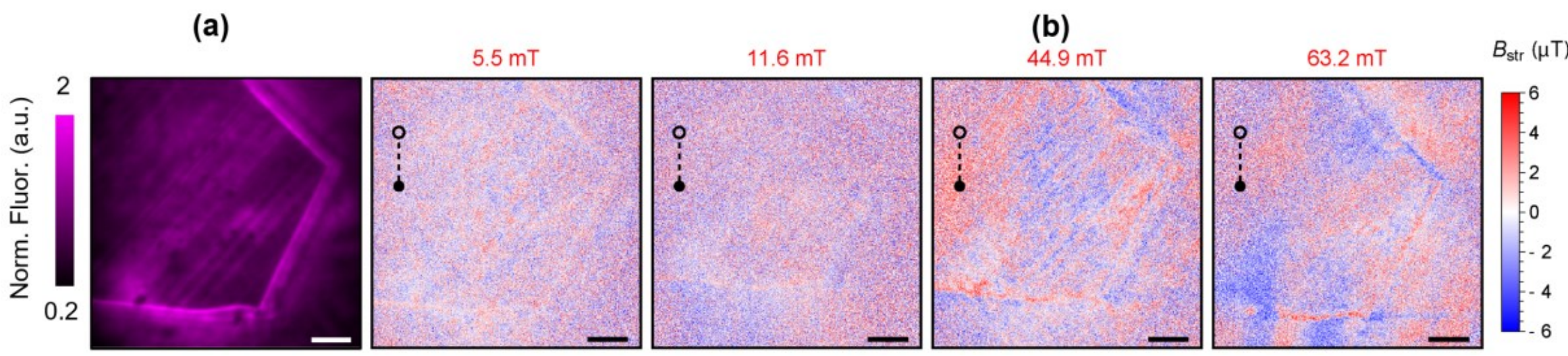

**Figure S3.3.** Normalized fluorescence map (a) and measured $B_{str}$ maps (b) at different values of $B_{app}$ (4.4 – 63.1 mT) for $WS_2$ Flake 3 transferred on (100) diamond. The scale bar in (a) and (b) is 5 μm.

Figure S3.4a and Figure S3.4b show the NV-measured and calculated $B_{str}$ maps of pristine $WS_2$ Flake 3 (thickness of ~45 nm), respectively. Figure S3.4c and Figure S3.4d depict line cut profiles of the measured $B_{str}$ at one (upper/lower) of Flake 1 (Figure S3.1) and Flake 3 edges at $B_{app}$ of 63.2 mT, as indicated by the dashed blue and red lines in (a) and (b). See the main text for the comparison discussion between Flake 1 and Falke 3.

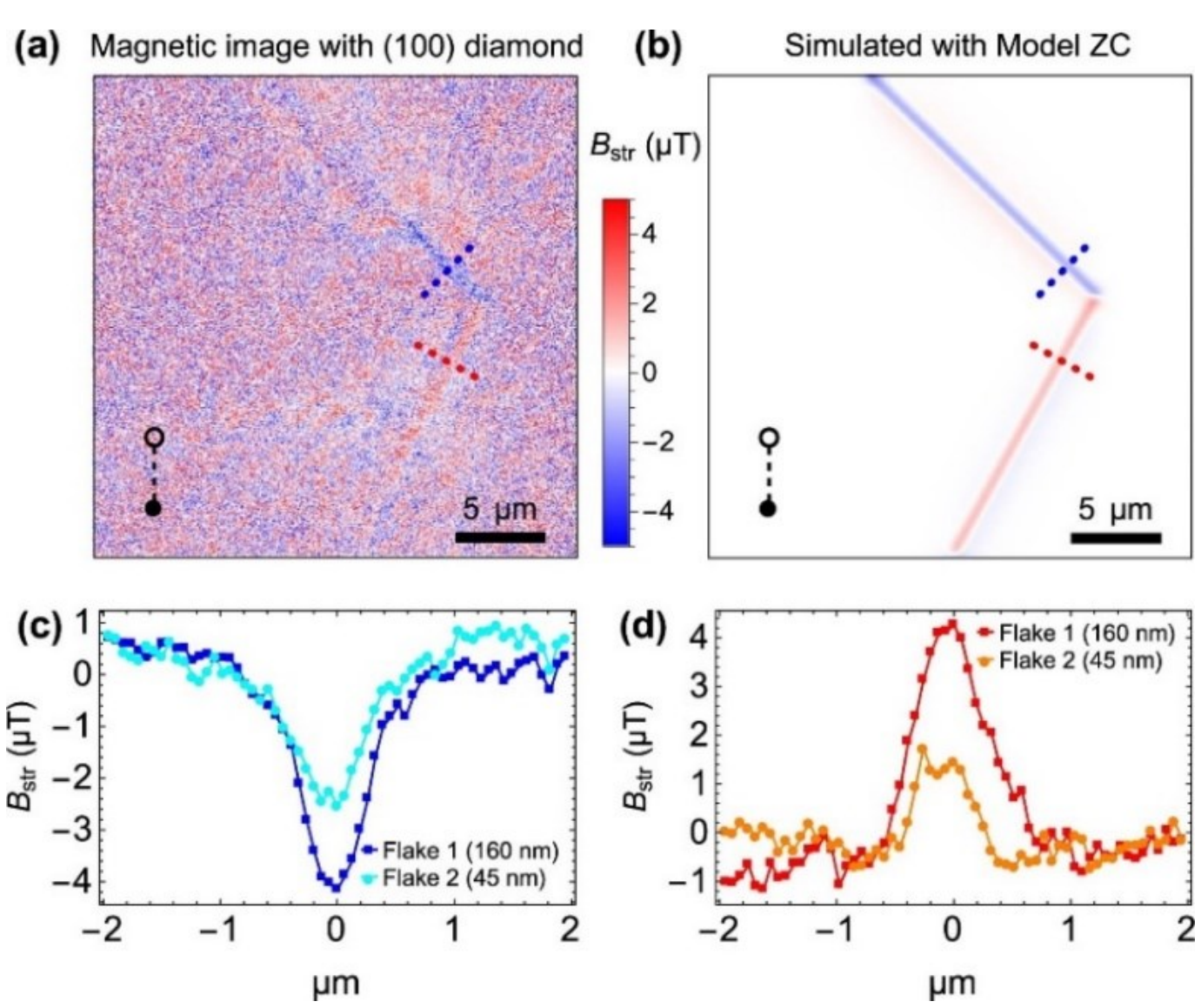

**Figure S3.4.** A measured (a) and simulated (b) $B_{str}$ map of $WS_2$ Flake 3 transferred on a (100) diamond at $B_{app}$ of 63.2 mT. Line cuts of $B_{str}$ across the edges of Flake 1 and Flake 3, corresponding to the blue (c) and red (d) curves for dashed lines in Figure 3b and Figure S3.4a, respectively, revealing field variations at different thicknesses.

### S4. COMSOL model

Figure S4 illustrates the schematic of edge magnetization for the $WS_2$ flake (Flake 3) used in our COMSOL model, as detailed in Section 2.1 of the main text. The magnetic stray field is integrated over a finite-thickness layer corresponding to the NV layer in the diamond substrate. For clarity, the applied magnetic field ($B_{app}$), aligned with the NV axis, is depicted but not directly included in the COMSOL model. The $WS_2$ flake's edges are modeled as bars with sub-resolution cross-sectional dimensions, and magnetization (50–80 A/m) is assigned to the volumes of these bars.

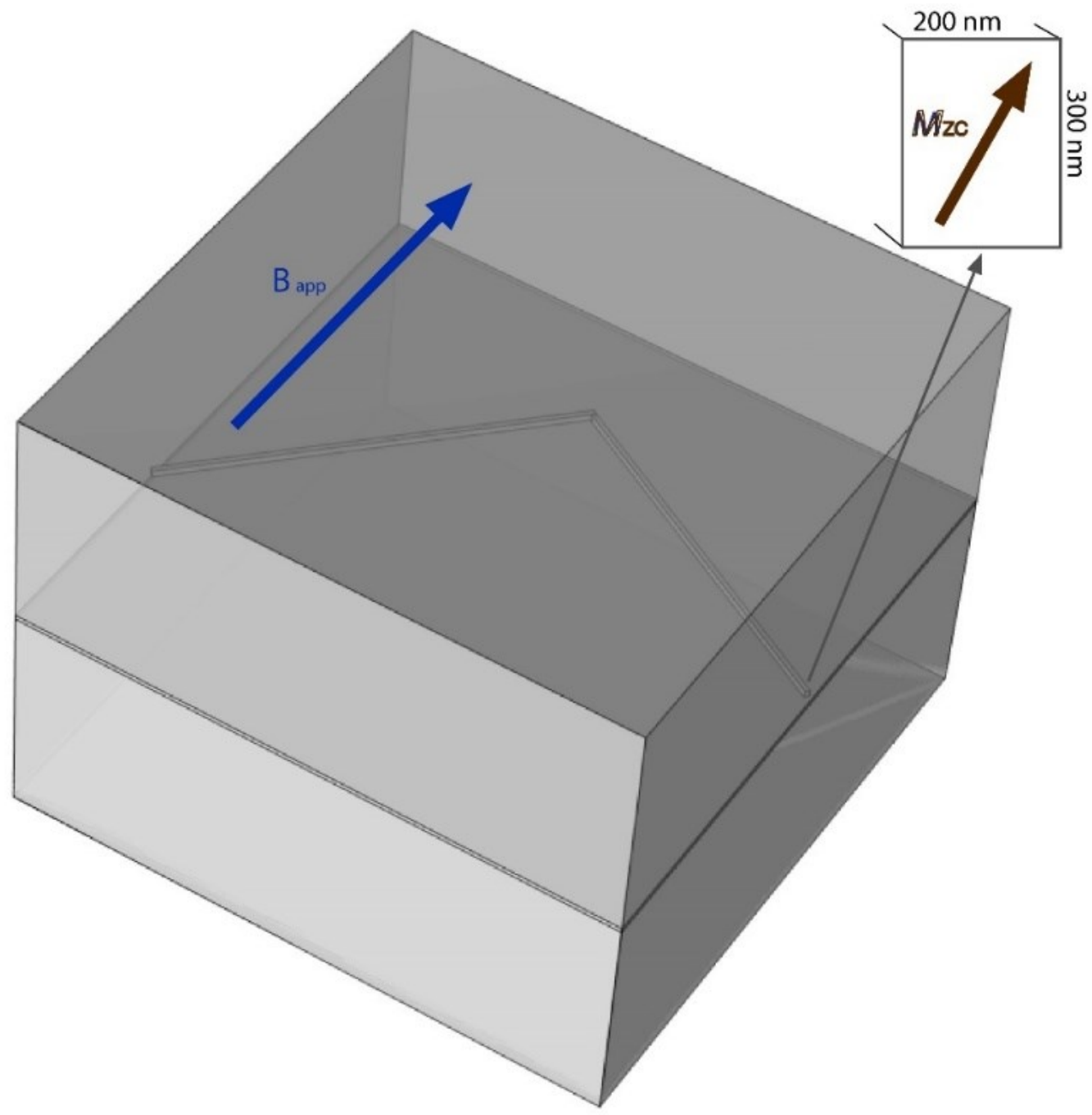

**Figure S4.** Schematic of the $WS_2$ flake (Flake 3) on a diamond substrate, illustrating the edge magnetization geometry used in the COMSOL model.